\documentclass[aps,prl,twocolumn,showpacs,superscriptaddress]{revtex4}

\usepackage{graphicx}

\begin{document}

\title{The boundary between long-range and short-range critical behavior}

\author{Erik Luijten}
\affiliation{Department of Materials Science and Engineering, University of
Illinois, Urbana, Illinois 61801}
\email{luijten@uiuc.edu}
%\homepage{http://ariadne.mse.uiuc.edu/}

\author{Henk W. J. Bl\"ote} 
\affiliation{Laboratory of Applied Physics, Delft University of Technology,
P.O. Box 5046, 2600 GA Delft, The Netherlands}
\affiliation{Lorentz Institute, Leiden University, P.O. Box 9506, 2300 RA
Leiden, The Netherlands}

\date{\today}

\begin{abstract}
  We investigate phase transitions of two-dimensional Ising models with
  power-law interactions, using an efficient Monte Carlo algorithm. For slow
  decay, the transition is of the mean-field type; for fast decay, it belongs
  to the short-range Ising universality class.  We focus on the intermediate
  range, where the critical exponents depend continuously on the power law. We
  find that the boundary with short-range critical behavior occurs for
  interactions depending on distance $r$ as $r^{-15/4}$. This answers a
  long-standing controversy between mutually conflicting renormalization-group
  analyses.
\end{abstract}


\maketitle

Many of the intermolecular forces that play a central role in large areas of
chemistry, physics, and biology have a long-ranged nature.  Well-known examples
are electrostatic interactions, polarization forces, and van~der~Waals forces.
Remarkably, there are still considerable deficiencies in our knowledge of the
critical behavior induced by these interactions, which include the prominent
case of Coulombic criticality~(see Ref.~\cite{mef-story} for a review).

But also for the simpler case of purely attractive, long-range interactions,
there exists a smothering controversy, which we aim to resolve in this work.
The present understanding of critical behavior in systems with such
algebraically decaying interactions is largely based on the
renormalization-group (RG) calculations for the O($n$) model by Fisher, Ma, and
Nickel~\cite{FMN}. Their analysis revealed that different regimes can be
identified for the universal critical properties, characterized by the decay
power of the interactions. In view of the small number of global parameters
determining the universality class, the location of the boundaries between
these regimes is of considerable interest. It is, therefore, disturbing to note
that there appears to be still no consensus on the theoretical side regarding
the precise location of the boundary between short-range and long-range
critical behavior. Here, we address this issue by means of numerical
calculations, which allow us, with minimal prior assumptions, to decide
between contradictory RG scenarios.

Our approach specializes to the Ising model, $n=1$, in $d$ dimensions,
described by the reduced Hamiltonian
\begin{equation}
\label{eq:hamil}
  \mathcal{H}/k_{\rm B}T = 
  - K \sum_{\langle ij\rangle} \frac{s_i s _j}{r_{ij}^{d+\sigma}} \;,
\end{equation}
where the spins $s_k=\pm 1$ are labeled by the lattice site $k$, the sum runs
over all spin pairs, and the pair coupling depends on the distance $r_{ij}=
|\vec{r}_i - \vec{r}_j|$ between the spins. According to the analysis of Fisher
\emph{et al.}~\cite{FMN}, universality classes are parametrized by $\sigma$,
and the following three distinct regimes were identified: a)~The classical
regime. The upper critical dimension is given by $d_u = 2 \sigma$, so that
mean-field-type critical behavior occurs for $\sigma \leq d/2$.  (b)~The
intermediate regime $d/2 < \sigma < 2$; here the critical exponents are
\emph{continuous functions} of $\sigma$. (c)~The short-range regime. For
$\sigma \geq 2$ the universal properties are those of the model with
short-range interactions, e.g., between nearest-neighbors only; thus one
observes that for $d=3$ van~der~Waals interactions (decaying as $1/r^6$)
actually lie quite close to the boundary between regimes (b) and~(c).

Although the general outline of these results has been widely accepted, one
part of this picture has become the scene of a debate that appears, up to now,
to be unresolved. This concerns the situation close to $\sigma = 2$.  In
Ref.~\cite{FMN} is has been conjectured that, throughout the intermediate
regime~(b), the correlation-function exponent~$\eta$ is \emph{exactly} given by
$2-\sigma$. On the other hand, in the short-range regime~(c), $\eta$ takes a
constant (but $d$-dependent) value $\eta_{\rm sr} > 0$ for all $d<4$, resulting
in a \emph{jump discontinuity} in~$\eta$ at the decay power $\sigma=2$. While
this remarkable phenomenon is not forbidden by thermodynamic arguments (which
only require $\eta \leq 2+\sigma$), it has attracted considerable attention
over the past decades and various efforts have been undertaken to reinvestigate
the corresponding RG scenario~\cite{Sak,Yama,GT,HN}. In addition, it may be
remarked that this scenario does not capture the one-dimensional case, where a
phase transition is rigorously known to be absent~\cite{Rue} for $\sigma >1$,
rather than for $\sigma > 2$ (see also Ref.~\cite{invsqr} and references
therein).  The issue was first addressed by Sak~\cite{Sak}, who pointed out
that higher-order terms in the RG equations considered in Ref.~\cite{FMN}
generate additional short-range interactions in the renormalization process,
which affects the competition between the long-range and the short-range fixed
points of the RG transformation.  As a consequence, for $d<4$ the boundary
between the intermediate and the short-range regime was found to shift from
$\sigma=2$ to $\sigma=2-\tilde\eta$, where the $\varepsilon$ expansion for
$\tilde\eta$ agrees to lowest orders with that of the short-range exponent
$\eta_{\rm sr}$.  In a field-theoretic approach, Honkonen and Nalimov~\cite{HN}
proved, to all orders in perturbation theory, the stability of the short-range
fixed point for $\sigma> 2-\eta_{\rm sr}$ and of its long-range counterpart for
$\sigma < 2-\eta_{\rm lr}$, where $\eta_{\rm lr}$ is the anomalous dimension of
the field, evaluated at the \emph{long-range} fixed point.  Note that these
authors also pointed out that the former result can be obtained from simple
scaling arguments, but the latter \emph{cannot}. Appealing aspects of these
findings are firstly the continuous and monotonic $\sigma$ dependence of the
correlation-function exponent (provided that $\eta_{\rm lr}$ and $\eta_{\rm
sr}$ coincide at $\sigma=2-\eta_{\rm sr}$), and secondly that the theory has
now attained consistency with the exact results for the one-dimensional case.

However, the analysis of~\cite{Sak} has also been the subject of criticism.
Van Enter~\cite{vE} pointed out an apparent inconsistency in Sak's results.
Three-dimensional \emph{short-range} O($n$) models exhibit a nonzero
spontaneous magnetization below the critical temperature. Van Enter~\cite{vE}
derived that, for $n \geq 2$, this broken symmetry is inconsistent with the
presence of a long-range perturbation with $\sigma < 2$. This contradicts Sak's
finding that such perturbations are irrelevant in a non-vanishing range
$2-\eta_{\rm sr} < \sigma <2$.  While this inconsistency emerges for $n\geq 2$
and antiferromagnetic perturbations only, it may be taken as a sign that the
renormalization scheme~\cite{Sak,HN} is incomplete, which could thus also
affect the purely ferromagnetic $n = 1$ case.  Another conflict with Sak's
results arose from the analyses of Yamazaki~\cite{Yama}.  A qualitative
difference between his and Sak's result for the correlation-function exponent
as a function of $\sigma$ is the absence of a kink at $\sigma=2-\eta_{\rm sr}$.
However, we note a problem with the internal consistency of his results: The
smoothly varying $\eta(\sigma)$ was obtained by using a long-range (i.e.,
$k^\sigma$-type, where $k$ is the wave vector used in the Fourier
representation) instead of a short-range ($k^2$-type) propagator in the RG
calculations, despite the fact that, in the same calculation, it was found that
the long-range term in the Landau-Ginzburg-Wilson Hamiltonian is irrelevant,
and only the short-range term survives.  A more serious objection was raised by
Gusm\~ao and Theumann~\cite{GT} (but ignored in Ref.~\cite{HN}), who argued
that the parameter $2-\sigma$ (i.e., $2\alpha$ in the notation of
Ref.~\cite{HN}) is \emph{not} a valid expansion parameter. They reconsidered
the problem using renormalized perturbation theory in $\varepsilon' =
2\sigma-d$, and found that the long-range expansion in $\varepsilon'$ is stable
with respect to short-range perturbations up to $\sigma=2$. This implies a
restoration of the early result of Fisher \emph{et al.}~\cite{FMN} that the
boundary between the intermediate and the short-range regime lies at $\sigma =
2$.  While, at least for $d=3$, the difference between this value and
$\sigma=2-\eta_{\rm sr}$ may be small, it indicates a fundamental dichotomy,
which is also relevant for other systems with competing fixed points.

Until recently, it appeared not to be feasible to shed some light on this
unsatisfactory situation by means of Monte Carlo simulations, for the following
reasons. Since the problems reside in the transition region between the
intermediate and the short-range regime, one expects corrections to scaling
to converge only slowly with increasing system size. This necessitates the
simulation of large systems. Since each spin interacts with all other spins in
the system, the amount of work per Metropolis-type spin visit then becomes
prohibitive. Use of a cluster Monte Carlo algorithm~\cite{LB95} that employs a
number of operations per spin flip that is \emph{independent of the system
size}, and moreover suppresses critical slowing down, has now enabled us to
obtain highly accurate data for sufficiently large system sizes.

We present numerical results for the critical exponent~$\eta$ and the Binder
cumulant~\cite{Binder} as a function of~$\sigma$, that allow us to distinguish
between the different theories. The simulated systems are defined on $L \times
L$ lattices with periodic boundaries and sizes from $L=4$ to $L=1000$. We
chose to study two-dimensional systems not only to maximize the attainable
linear system size, but in particular because there the exponent $\eta_{\rm
sr}=\frac{1}{4}$ has a much larger value than for $d=3$ ($\eta_{\rm sr} \simeq
0.037$); this maximizes both the size of the disputed region $\langle
2-\eta_{\rm sr}, 2 \rangle$ and the magnitude of the supposed jump in
$\eta(\sigma)$.  The length of the simulations was such, that (for the largest
system sizes) typically a relative uncertainty of one part in a thousand was
reached for the Binder cumulant.  The precise form of the pair interaction was
taken as
\begin{equation}
\label{eq:interac}
\tilde{K} (|\vec{r}|)= K
            \int_{r_x-\frac{1}{2}}^{r_x+\frac{1}{2}} {\rm d}x
            \int_{r_y-\frac{1}{2}}^{r_y+\frac{1}{2}} {\rm d}y
            \frac{1}{(x^2 + y^2)^{(d+\sigma)/2}} \;,
\end{equation}
where $\vec{r}=(r_x,r_y)$ denotes the difference between the integer
coordinates of the two interacting spins. It is important to note that this
differs from the interaction in Eq.~(\ref{eq:hamil}) only in powers of $r$ that
decay faster than $r^{-d-\sigma}$; these terms are irrelevant and will only
affect nonuniversal terms, such as the location of the critical point. The
critical exponents and boundaries of the regimes (a)--(c) will not be altered.
As mentioned, we have adopted periodic boundary conditions, which are expected
to speed up the approach to the thermodynamic limit. No artificial cutoff was
imposed on the interaction range: Interactions take place between all periodic
images.

We have carried out simulations for decay powers $\sigma=1.2$, $1.4$, $1.6$,
$1.75$, $1.85$, $1.95$, $2.0$, $2.05$, $2.25$, $2.5$, $2.75$, and $3.0$. For
each value of $\sigma$, we have determined the critical coupling $K_c$ on
the basis of a finite-size scaling analysis of the susceptibility-like quantity
$\langle m^2\rangle$ and of the dimensionless amplitude ratio $Q \equiv \langle
m^2 \rangle^2 / \langle m^4 \rangle$, where $m$ is the magnetization density.
$Q$ is a variant of the fourth-order cumulant introduced by
Binder~\cite{Binder}.  For Ising-like systems above the upper critical
dimension, i.e., in the classical regime, it has been predicted~\cite{brez-zj}
to take the universal critical value $8\pi^2/\Gamma^4(\frac{1}{4}) \simeq
0.456947$, the same value as in the mean-field model~\cite{LB95}; this has been
confirmed numerically in Ref.~\cite{LB9697}.  In the short-range regime it
takes a universal value close to $Q=0.856216\cdots$ \cite{KB,SS}.  In the
intermediate regime, finally, $Q$ has been calculated by means of a singular
expansion in $\varepsilon'$, up to second order in
$\sqrt{\varepsilon'}$~\cite{EL99,CT01}.  From the magnetic susceptibility,
which diverges as $L^{\gamma/\nu}$ at criticality, $\eta$ can be extracted
using the scaling law $\gamma = (2-\eta) \nu $.  An important issue in these
finite-size analyses is the fact that the correction-to-scaling exponents are
essentially unknown, and actually depend on the borderline value of~$\sigma$.
In particular, there will be logarithmically decaying finite-size corrections
at the borderline itself. Considerable effort has been exercised to encompass
all such uncertainties in the quoted margins for our estimates for $K_c$,
$Q$, and $\eta$ (see Table~\ref{tab:critical}).

\begin{table}
\caption{Results of detailed finite-size-scaling analyses for two-dimensional
  systems with power-law interactions, for several values of the
  decay parameter $\sigma$. The results for the critical point $K_c$,
  the fourth-order amplitude ratio $Q$ and the correlation-function
  exponent~$\eta$ are followed by the estimated error in the last decimal
  places. The last column lists the predicted values for $\eta$, see the
  text.}
\label{tab:critical}
\label{tab:qkc}
\begin{tabular}{l|l|l|l|l}
 $\sigma$ & $K_c$          & $Q$        & $\eta$      & $\eta$ [predicted] 
 \\ \hline
 1.20     &  0.114966 (3)  & 0.559 (4)  & 0.798 (18)  & 0.80        \\
 1.40     &  0.125300 (4)  & 0.650 (3)  & 0.616 (10)  & 0.60        \\
 1.60     &  0.133397 (5)  & 0.752 (7)  & 0.410 (24)  & 0.40        \\
 1.75     &  0.137872 (2)  & 0.840 (10) & 0.286 (24)  & 0.25        \\
 1.85     &  0.140073 (3)  & 0.85  (3)  & 0.30  (6)   & 0.25 \text{or} 0.15  \\
 1.95     &  0.141644 (3)  & 0.849 (14) & 0.24  (4)   & 0.25 \text{or} 0.05  \\
 2.00     &  0.142198 (3)  & 0.850 (6)  & 0.266 (16)  & 0.25        \\
 2.05     &  0.142610 (8)  & 0.857 (7)  & 0.260 (14)  & 0.25        \\
 2.25     &  0.142831 (9)  & 0.856 (7)  & 0.248 (12)  & 0.25        \\
 2.50     &  0.140401 (8)  & 0.860 (3)  & 0.246 (8)   & 0.25        \\
 2.75     &  0.135636 (11) & 0.855 (3)  & 0.250 (10)  & 0.25        \\
 3.00     &  0.129267 (12) & 0.857 (3)  & 0.248 (8)   & 0.25
\end{tabular}
\end{table}

While the detailed least-squares analysis depended on the value of~$\sigma$,
the following general procedure was applied. An (effective) leading
correction-to-scaling exponent~$y_i$ was obtained from a fit of the amplitude
ratio~$Q$, in which also temperature-dependent terms were taken into account.
Subsequently, $\gamma/\nu$ was extracted from an analysis of $\langle m^2
\rangle$ in which $y_i$ was kept fixed. This allowed us to include higher-order
corrections. The analysis was then repeated with $y_i$ varied within its
uncertainty margins, in order to determine the uncertainty in $\gamma/\nu$.
Finally, the entire analysis was repeated for different minimum system sizes.
For $\sigma=1.75$, a consistent analysis could only be obtained by allowing for
logarithmic corrections to scaling, as would be expected if the crossover
from long-range to short-range criticality occurs at $\sigma=2-\eta_{\rm sr}$.

The critical coupling exhibits a very smooth but also remarkable, almost
parabola-shaped dependence on $\sigma$, with a maximum around $\sigma=2.17$.
This nonmonotonic variation may be due to our use of the integrated
coupling~(\ref{eq:interac}) which tends to enhance the close-neighbor
couplings, in particular for large $\sigma$. For $\sigma>2$, both $Q$ and
$\eta$ are fully consistent with the known values for the two-dimensional Ising
universality class. On the other hand, for $\sigma \leq 1.60$, $\eta$ is in
excellent agreement with the RG prediction $2-\sigma$~\cite{FMN}; $Q$ exhibits
a remarkably linear deviation from its mean-field value~\cite{EL99}.  However,
in the disputed region $1.75 \leq \sigma < 2.0$, where we invested most of our
computing efforts, the uncertainties are still considerable; this is by itself
already indicative of the occurrence of a crossover between two fixed points.

\begin{figure}
\leavevmode\centering
\includegraphics[width=85mm]{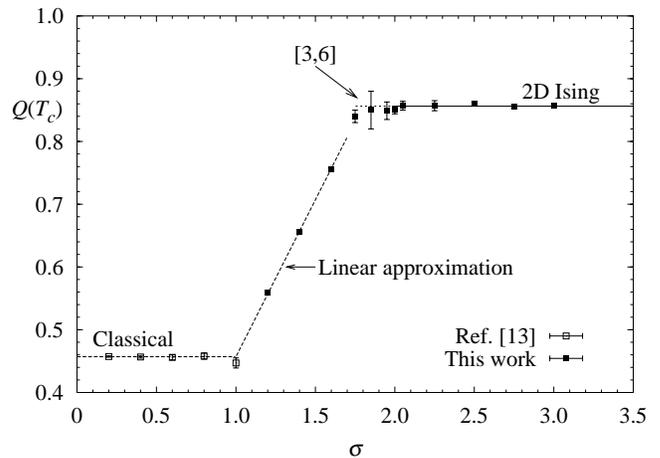}
\caption{The amplitude ratio $Q$ as a function of the decay
parameter~$\sigma$. The behavior is consistent with a crossover from long-range
to short-range critical behavior around $\sigma=1.75$, as predicted in
Refs.~\protect\cite{Sak,HN}.}
\label{fig:Q}
\end{figure}

In the absence of accurate predictions for the amplitude ratio $Q$ (depicted in
Fig.~\ref{fig:Q}), this quantity does not provide much further evidence to
distinguish between the different theoretical scenarios. However, there is no
evidence for a jump in $Q$ at $\sigma=2$; this fails to support the Gusm\~ao
and Theumann~\cite{GT} analysis. For $\sigma=1.85$ and $\sigma=1.95$, we find
that $Q$ is compatible with the 2D Ising value, and the deviation of about
1.5 standard errors at $\sigma=1.75$ is still within acceptable limits. 

\begin{figure}
\leavevmode\centering
\includegraphics[width=85mm]{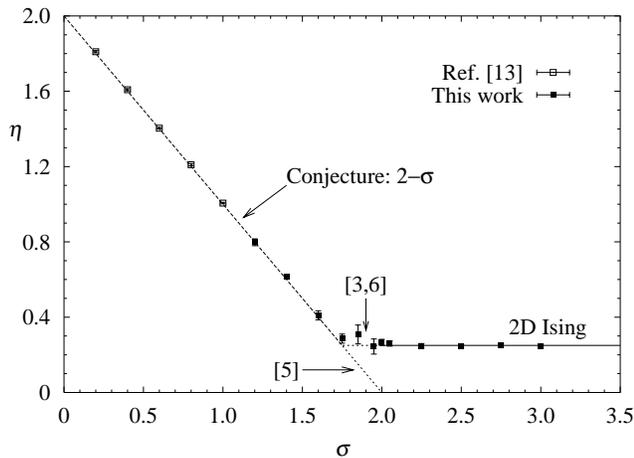}
\caption{The correlation-function exponent $\eta$ as a function of the decay
parameter~$\sigma$. For $\sigma<1.75$ the data are in excellent agreement with
the conjecture $\eta=2-\sigma$~\protect\cite{FMN}, whereas for $\sigma>1.75$
they agree with the $d=2$ short-range Ising universality class. A
crossover at $\sigma=2$~\protect\cite{GT} is clearly excluded.}
\label{fig:eta}
\end{figure}

On the other hand, the exponent~$\eta$ offers clear evidence in order to
distinguish between the different scenarios, cf.\ Fig.~\ref{fig:eta}. While
also this quantity suffers from relatively large uncertainties in the crossover
region, these uncertainties are small enough to exclude the scenario of Ref.
\cite{GT} which implies that a transition to short-range critical behavior
occurs at $\sigma=2$ and that $\eta = 2 - \sigma$ for $\sigma<2$. Indeed,
instead of the 2D Ising value $\eta = \frac{1}{4}$, this exponent should then
take the values $0.10$ at $\sigma=1.85$ and $0.05$ at $\sigma=1.95$.  However,
at these two values of $\sigma$, our results for $\eta$ lie within one standard
deviation from the 2D Ising value, and three respectively almost five standard
deviations above $2-\sigma$.

In summary, we have demonstrated that both the correlation-function exponent
$\eta$ and the fourth-order amplitude ratio $Q$ take their (universal)
short-range Ising values for $\sigma > 2-\eta_{\rm sr}$. For $\sigma>2$ this
could be shown with high numerical accuracy, but also for $2-\eta_{\rm sr} <
\sigma < 2$ the results are precise enough to convincingly exclude a crossover
from long-range to short-range critical behavior at $\sigma=2$. Instead, they
support a crossover at $\sigma=2-\eta_{\rm sr}$, in agreement with
Refs.~\cite{Sak,HN}.  This provides a clear answer to the dilemma posed by
conflicting renormalization-group scenarios.

Furthermore, our findings for $\eta$ are in excellent agreement with
$\eta=2-\sigma$ in the intermediate range $d/2 < \sigma < 2-\eta_{\rm sr}$,
supporting the conjecture that all higher-order contributions vanish in the
$\varepsilon'$ expansion for $\eta$~\cite{FMN} and in contrast with the
scenario proposed in Ref.~\cite{Yama}.

The amplitude ratio $Q$ depends approximately linearly on $\sigma$ for $d/2 <
\sigma < 2-\eta_{sr}$. Remarkably, the most prominent deviations from linearity
occur near $\sigma = 2-\eta_{sr}$, while the $\varepsilon'$ expansion predicts
a square-root-like singularity at the opposite end of the intermediate
range~\cite{EL99}.

Finally, we quote a plausible explanation for the breakdown of the description
of Ref.~\cite{GT} in the range $2-\eta_{\rm sr}<\sigma<2$~\cite{Sakpc}.  If the
validity of the $\varepsilon'$-expansion is not uniform in $\sigma$, the region
of convergence may shrink to zero when $\sigma$ approaches~2.  Then one cannot
fix $\varepsilon'$ and take the limit $\sigma\rightarrow 2$, as is done in
Ref.~\cite{GT}.  This would also explain why the long-range ($\varepsilon'$)
expansion does not provide any indication of a change to short-range behavior
at $\sigma=2-\eta_{\rm sr}$.

\begin{acknowledgments}
  We are indebted to Professor J. Sak for providing his valuable comments on
  the status of the theoretical analyses of the border between the regions of
  the intermediate- and the short-range interactions.  This research is
  supported in part by the Dutch FOM foundation (``Stichting voor Fundamenteel
  Onderzoek der Materie'') which is financially supported by the NWO
  (``Nederlandse Organisatie voor Wetenschappelijk Onderzoek'').
\end{acknowledgments}

\end{document}